\documentstyle[psfig]{mn}

\newif\ifAMStwofonts
\def\lapp{\ifmmode\stackrel{<}{_{\sim}}\else$\stackrel{<}{_{\sim}}$\fi}
\def\gapp{\ifmmode\stackrel{>}{_{\sim}}\else$\stackrel{>}{_{\sim}}$\fi}

\title[Evidence for alignment of the rotation and velocity vectors
in pulsars]
{Evidence for alignment of the rotation and velocity vectors
in pulsars}
\author[Johnston et al.]
{Simon Johnston$^1$, G. Hobbs$^1$, S. Vigeland$^2$, 
M. Kramer$^3$, \newauthor J. M. Weisberg$^{1,4,2}$ \& A. G. Lyne$^3$\\
$^1$Australia Telescope National Facility, CSIRO, P.O. Box 76, 
Epping, NSW 1710, Australia. \\
$^2$ Dept. of Physics and Astronomy, Carleton College, Northfield, MN 55057, USA.\\
$^3$ University of Manchester, Jodrell Bank Observatory, Macclesfield, Cheshire, SK11 9DL.\\
$^4$ School of Physics, University of Sydney, NSW 2006, Australia.
}
\date{\today}
\pagerange{\pageref{firstpage}--\pageref{lastpage}}
\pubyear{2005}
\begin{document}
\maketitle
\label{firstpage}

\begin{abstract}
We present strong observational evidence for a 
relationship between the direction of a pulsar's motion and its
rotation axis. We show carefully calibrated polarization data for 25
pulsars, 20 of which display linearly polarized emission from the pulse
longitude at closest approach to the magnetic pole. Such data allow
determination of the position angle of the linear polarisation which
in turn reflects the position angle of the rotation axis. Of
these 20 pulsars, 10 show an offset between the velocity vector and
the polarisation position angle which is either less than 10\degr\ or
more than 80\degr, a fraction which is very unlikely by random
chance. We believe that the bimodal nature of the distribution arises from the
presence of orthogonal polarisation modes in the pulsar radio emission.
In some cases this orthogonal ambiguity
is resolved by observations at other wavelengths so that
we conclude that the velocity vector and the rotation axis are
aligned at birth. Strengthening the case is
the fact that 4 of the 5 pulsars with ages less than 3~Myr show this
relationship, including the Vela pulsar.
We discuss the implications of these
findings in the context of the Spruit \& Phinney (1998)\nocite{sp98}
model of pulsar birth-kicks.
We point out that, contrary to claims in the
literature, observations of double neutron star systems do not rule
out aligned kick models and describe a possible observational test
involving the double pulsar system.
\end{abstract}


\section{Introduction}
The velocities of pulsars are significantly larger than those of their
progenitor (high-mass) stars (e.g. Lyne \& Lorimer
1994)\nocite{ll94}. This implies that the birth process of pulsars
also produces their high velocities and that the supernova or events
soon thereafter must be asymmetrical.  The mechanisms driving this
asymmetry are far from clear.  Tademaru \& Harrison
(1975)\nocite{th75} proposed the so-called rocket mechanism whereby an
offset of the magnetic dipole from the centre of the star causes the
star to accelerate along its rotation axis. Observational tests of
this mechanism by Morris, Radhakrishnan \& Shukre (1976)\nocite{mrs76}
and Tademaru (1977)\nocite{tad77} reached opposing conclusions, with
the former claiming no correlation between the velocity vector and the
spin axis, and the latter claiming a good correlation.  Following
this, Anderson \& Lyne (1983)\nocite{al83} revisited the issue with an
updated database of pulsar proper motions. They also saw no apparent
alignment between the proper motion angle and the rotation axis
although they included many old pulsars in their sample, whose
direction of motion may have significantly evolved in the Galactic
potential.  In the late 1980s, a group of papers concentrated mainly
on asymmetric explosions and the break-up of binary systems as the
origin for pulsar velocities \cite{rs86,dc87,bai89} and the rocket
model fell out of favour.

A decade later, the debate started anew when Spruit \& Phinney
(1998)\nocite{sp98} and Cowsik (1998)\nocite{cow98} proposed a model
in which the same (off-centre) explosion processes that give pulsars
their high velocity also produce their fast rotation.  Although their
models were motivated by theoretical results suggesting that the
pre-supernova stellar configuration is not spinning quickly enough to
account for birth periods of $\sim$10~ms from conservation of angular
momentum alone, the basic physics applies - any significant off-centre
impulses would both accelerate {\em and} torque the pulsar.  In the
Spruit \& Phinney (1998)\nocite{sp98} model, a single, short-duration
impulse leads to the rotation axis being perpendicular to the velocity
vector. However, even a few short duration kicks as the star spins
serve to destroy this correlation. They also considered kicks of
finite duration ($\geq$ the spin period) and showed in this case that
the rotation axis should be parallel with the velocity vector.
Deshpande, Ramachandran \& Radhakrishnan (1999)\nocite{drr99} tested
these ideas by using the literature to search for pulsars which had
accurate proper motion measurements and good polarization
observations. They used the polarization measurements to determine the
point of inflexion of the PA sweep although give no details as to how
this was done.  They took the associated position angle as the angle
of the rotation axis on the sky (or, to allow for orthogonal mode
emission, its orthogonal).  In their sample of 29 pulsars they saw no
statistical evidence for any correlation between the velocity vector
and the rotation axis. In the context of the Spruit \& Phinney (1998)
model they favoured multiple short duration impulses as the origin of
the spin and velocity kicks.

Observational evidence {\it for} a correlation between the spin axis
and the proper motion next came from an unexpected source.  Helfand et
al. (2001)\nocite{hgh01} showed high resolution X-ray images of the
Vela pulsar and its surrounding nebula. Two prominent arcs are seen,
which Helfand et al. (2001) interpret as originating from the front
surface of an equatorial wind whose axis of symmetry is the rotation
axis of the pulsar. The position angle of the axis of this torus is
130\degr\ and 310\degr.  The axis of the torus, created by pulsar
winds, is likely aligned with the pulsar rotation axis.  Helfand et
al. (2001) then note that the position angles of the torus and pulsar
spin axes are within 10\degr\ of the proper motion vector of 301\degr\
obtained by Dodson et al. (2003)\nocite{dlrm03}.  Following from the
Vela result, X-ray imaging of a number of young pulsars showed that
many had pulsar wind nebulae consisting of an equatorial torus and
polar jets.  Ng \& Romani (2004)\nocite{nr04} used this information to
determine the angle of rotation axis on the plane of they sky and
compared this to the velocity vector if available. For the Crab and
Vela pulsars these clearly matched but there was also good evidence
for a match in the other pulsars in their sample
(PSRs J0538+2817, B1951+32 and B1706--44). It appears
likely from this study that the directions of motion of young pulsars
lie along the rotation axes, although their sample is small.

Historically, the velocity vectors of pulsars have been measured
either through proper motion measurements using Very Long Baseline
Interferometry (VLBI) techniques or through highly accurate timing.
The first method tends to be possible only for relatively nearby
pulsars, whereas the second method was generally restricted to
millisecond pulsars which show almost no intrinsic timing jitter.
Recently, however, Hobbs et al. (2004)\nocite{hlk+04} have shown that
it is possible to measure proper motions of pulsars even in the
presence of timing noise. In brief, this involves very long (tens of
years) timing sets coupled with a `whitening' technique to remove the
timing noise. Using this method, Hobbs et al. (2004)\nocite{hlk+04}
list proper motion measurements for 302 pulsars. Updates to these
parameters and a catalogue of the best available proper motions
from either timing or interferometry are available in
Hobbs et al. (2005)\nocite{hllk05} and in particular provide celestial 
position angles (PA$_v$) of the velocity vectors.

In the absence of external features such as pulsar wind nebulae and
tori, the position angle of the axis of rotation (PA$_r$) of a pulsar
has to be determined from polarisation observations using a
several-step procedure.  One first has to determine PA$_0$, the
position angle of polarization at the point of closest approach of the
observer's line of sight to the magnetic pole. This angle must then be
corrected for Faraday rotation in the interstellar medium and the
ionosphere in order to reflect the true angle at the pulsar.
Determination of PA$_0$ itself is difficult. We discuss two
techniques.

In the rotating vector model (RVM) of Radhakrishnan \& Cooke
(1969)\nocite{rc69a}, the radiation is beamed along the field lines
and the plane of polarization is determined by the angle of the
magnetic field as it sweeps past the line of sight. The PA as a
function of pulse longitude, $\phi$, can be expressed as
\begin{equation}
{\rm PA} = {\rm PA}_{0} + 
{\rm arctan} \left( \frac{{\rm sin}\alpha 
 \, {\rm sin}(\phi - \phi_0)}{{\rm sin}\zeta
 \, {\rm cos}\alpha - {\rm cos}\zeta
 \, {\rm sin}\alpha \, {\rm cos}(\phi - \phi_0)} \right)
\end{equation}
Here, $\alpha$ is the angle between the rotation axis and the magnetic
axis, where $\zeta=\alpha+\beta$, $\beta$ being the angle at closest approach
of the line of sight to the magnetic axis. 
$\phi_0$ is the corresponding pulse longitude at which the PA is
then PA$_{0}$.  Unfortunately, for most pulsars it is difficult to
determine $\alpha$ and $\beta$ with any degree of accuracy, partly
because the longitude over which pulsars emit is rather small and
partly because strong deviations from a simple swing of PA are often
observed.  This makes the determination of PA$_0$ straightforward only
in the $\sim$15 per cent of pulsars for which the RVM works (see also
Everett \& Weisberg 2001\nocite{ew01}).

\begin{figure}
\centerline{\psfig{figure=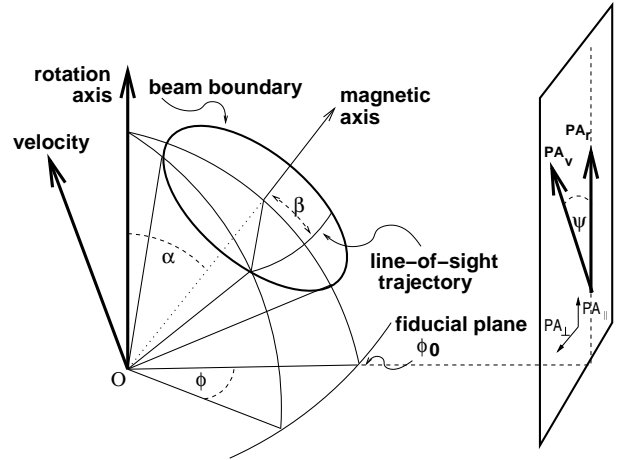,angle=0,width=8cm}}
\caption{ 
Geometry of the pulsar beam and velocity vector as seen by the
observer. The observer sees a projection of the direction of the
rotation axis, PA$_r$, and the velocity vector, PA$_v$,
onto the plane-of-the sky shown on the right.
The angle $\alpha$ is the inclination angle
of the magnetic and rotation axis, while $\beta$ is the impact
parameter between the magnetic axis and the line-of-sight at
closest approach.
The vector PA$_\parallel$ lies along the field line
in this fiducial plane ($\phi_0$) and hence
has the same value as PA$_r$. PA$_\perp$ lies orthogonal to the magnetic field
at closest approach. In the absence of other information we do not know
whether the observed PA$_0$ corresponds to PA$_\parallel$ or PA$_\perp$.
Note that PA$_\parallel$ and PA$_\perp$ (and hence PA$_r$)
have 180\degr\ ambiguities inherent to all polarization measurements,
unlike PA$_v$.}
\label{all}
\end{figure}
\begin{table*}
\caption{Proper motions, rotation measures and polarisation position
angles for a sample of 25 pulsars.}
\setlength{\tabcolsep}{3pt}
\begin{tabular}{llclrrrrrrrr}
\hline & \vspace{-3mm} \\
& & \multicolumn{1}{c}{Age} & \multicolumn{1}{c}{Dist}
& \multicolumn{4}{c}{Proper Motion}
& \multicolumn{3}{c}{Rotation Measure} & \multicolumn{1}{c}{Poln} \\
\multicolumn{1}{c}{Jname} & \multicolumn{1}{c}{Bname} 
& \multicolumn{1}{c}{log($\tau_c$)} &
& \multicolumn{1}{c}{$\mu_\alpha$} & \multicolumn{1}{c}{$\mu_\delta$}
& \multicolumn{1}{c}{Ref} & \multicolumn{1}{c}{PA$_v$} 
& \multicolumn{1}{c}{Previous} & \multicolumn{1}{c}{Ref}
& \multicolumn{1}{c}{Measured} & \multicolumn{1}{c}{PA$_0$} \\
& & \multicolumn{1}{c}{(yr)} & \multicolumn{1}{c}{(kpc)} 
&\multicolumn{1}{c}{(mas/yr)}
& \multicolumn{1}{c}{(mas/yr)} & & \multicolumn{1}{c}{(deg)}
& \multicolumn{1}{c}{(rad~m$^{-2}$)} &
& \multicolumn{1}{c}{(rad~m$^{-2}$)} 
& \multicolumn{1}{c}{(deg)} \\
\hline & \vspace{-3mm} \\
J0525+1115   & B0523+11   & 7.9 & 3.1 & 27(2)       & $-$24(10)   & 1 & 132(16)   & 35(3) & 1 & 37(2) & $-$65(4)\\
J0601$-$0527 & B0559$-$05 & 6.7 & 3.9 & $-$5(3)     & $-$20(6)    & 1 & 194(16)   & 67(5) & 2 & 62.4(20) & --- \\
J0630$-$2834 & B0628$-$28 & 6.4 & 1.4 & $-$44.6(9)  & 19.5(22)    & 2 & 294(3)    & 46.19(9) & 3 & 46.53(12) & 26(2) \\
J0742$-$2822 & B0740$-$28 & 5.2 & 2.1 & $-$29(2)    & 4(2)        & 3 & 278(5)    & 156(5) & 4 & 149.95(5)& $-$81.7(1)\\
J0820$-$1350 & B0818$-$13 & 7.0 & 2.0 & 18(3)       & $-$46(5)    & 1 & 159(6)    & $-$1.2(4) & 3 & $-$7.2(12)& 65(2)\\
\vspace{-2mm} \\
J0835$-$4510 & B0833$-$45 & 4.1 & 0.29 & $-$49.60(6) & 29.8(1)     & 4 & 301.0(1)  & 38.17(6) & 5 & 31.38(1) & 36.8(1)\\
J0908$-$1739 & B0906$-$17 & 7.0 & 0.9 & 22(3)       & $-$94(4)    & 1 & 167(2) & $-$36(6) & 2 & $-$31(4) & --- \\
J0922+0638   & B0919+06   & 5.7 & 1.2 & 18.35(6)    & 86.56(12)   & 5 &  12.0(1)  & 32(2) & 2 & 29.2(3) & --- \\
J0953+0755   & B0950+08   & 7.2 & 0.26 & $-$2.09(8)  & 29.46(7)    & 6 & 355.9(2)  & 1.35(15) & 3 & $-$0.66(4) & 14.9(1)\\
J1136+1551   & B1133+16   & 6.7 & 0.36 & $-$74.0(4)  & 368.1(3)    & 6 & 348.6(1)  & 3.9(2) & 6 & 1.1(2) & $-$78(2)\\
\vspace{-2mm} \\
J1239+2453   & B1237+25   & 7.4 & 0.86 & $-$106.82(17) & 49.92(18) & 6 & 295.0(1)  & $-$0.33(6) & 3 & $-$2.6(4) & $-$66(1)\\
J1430$-$6623 & B1426$-$66 & 6.7 & 1.0 & $-$31(5)    & $-$21(3)    & 7 & 236(9)    & $-$12(3) & 3 & $-$19.2(3) & $-$28.5(7)\\
J1453$-$6413 & B1449$-$64 & 6.0 & 2.1 & $-$16(1)    & $-$21.3(8)  & 7 & 217(3)    & $-$22.3(14) & 3 & $-$18.6(2) & $-$56.9(4)\\
J1456$-$6843 & B1451$-$68 & 7.6 & 0.45 & $-$39.5(4)  & $-$12.3(3)  & 8 & 252.7(6)  & $-$2(3) &  3 & $-$4.0(3) & $-$31.6(6)\\
J1645$-$0317 & B1642$-$03 & 6.5 & 1.1 & $-$3.7(15)  & 30.0(16)    & 2 & 353(3)    & 15.8(3) & 3 & 17(2) & 56(4)\\
\vspace{-2mm} \\
J1709$-$1640 & B1706$-$16 & 6.2 & 0.8 & $-$9(3)     & $-$44(21)   & 1 & 192(16)   & $-$1.3(3) & 2 & $-$2(5) & 15(2)\\
J1709$-$4429 & B1706$-$44 & 4.2 & 2.3 &             &             &   & 160(10)   & $-$7(4) & 7 & 0.70(7) & --- \\
J1740+1311   & B1737+13   & 6.9 & 1.5 & $-$21.5(22) & $-$19.7(22) & 2 & 228(6)    & 64.4(16) & 1 & 65(2) & $-$46(4)\\
J1844+1454   & B1842+14   & 6.5 & 2.2 & 18(4)       & 25(7)       & 1 & 36(15)    & 121(8) & 2 & 109.0(13) & $-$52(2)\\
J1900$-$2600 & B1857$-$26 & 7.7 & 2.0 & $-$19.9(3)  & $-$47.3(9)  & 9 & 202.8(7)  & $-$7.3(8) & 3 & $-$2.3(8) & $-$43(2)\\
\vspace{-2mm} \\
J1913$-$0440 & B1911$-$04 & 6.5 & 2.8 & 6(2)        & $-$24(7)    & 1 & 166(11)   & 12(3) & 2 & 4.4(9) & $-$68(2)\\
J1921+2153   & B1919+21   & 7.2 & 1.1 & 19(4)       & 28(6)       & 1 & 34(12)    & $-$16.5(5) & 3 & $-$13(3) & $-$35(2)\\
J1932+1059   & B1929+10   & 6.5 & 0.36 & 94.03(14)   & 43.4(3)     & 5 & 65.2(2)   & $-$6.1(1.0) & 3 & $-$6.87(2) & $-$11.3(1)\\
J1935+1616   & B1933+16   & 6.0 & 5.6 & 1.1(4)      & $-$17.6(7)  & 1 & 176.4(14) & $-$1.9(4) & 6 & $-$10.2(3) & 10.1(7)\\
J1941$-$2602 & B1937$-$26 & 6.8 & 1.7 & 12(2)       & $-$10(4)    & 2 & 130(17)   & $-$41(4) & 7 & $-$33.5(8) & --- \\
\hline & \vspace{-3mm} \\
\end{tabular}

Proper motion references:
1. updated parameters since Hobbs et al. (2004)\nocite{hlk+04},
2. Brisken et al. (2003)\nocite{bfg+03},
3. Fomalont et al. (1997)\nocite{fgml97},
4. Dodson et al. (2003)\nocite{dlrm03},
5. Chatterjee et al. (2001)\nocite{ccl+01},
6. Brisken et al. (2002)\nocite{bbgt02},
7. Bailes et al. (1990b)\nocite{bmk+90a},
8. Bailes et al. (1990a)\nocite{bmk+90b},
9. Fomalont et al. (1999)\nocite{fgbc99}.
Rotation measure references:
1. Weisberg et al. (2004)\nocite{wck+04},
2. Hamilton \& Lyne (1987)\nocite{hl87},
3. Hamilton, McCulloch \& Manchester (1981; unpublished),
4. Han, Manchester \& Qiao (1999)\nocite{hmq99},
5. Hamilton et al. (1987)\nocite{hmm+77},
6. Manchester (1972)\nocite{man72},
7. Qiao et al. (1995)\nocite{qmlg95}
\label{sources}
\end{table*}

It is possible, however, to determine $\phi_0$ and hence PA$_0$ without
RVM fitting. The frequency evolution of pulsar profiles has shown
that components close to $\phi_0$ have steeper spectra than components
at larger values of $\phi$ \cite{ran83,lm88,kra94}.
Also, there is often change in the
handedness of circular polarization at $\phi_0$ \cite{rr90}. Finally, symmetry
in pulse profiles is also an important
indicator of $\phi_0$ \cite{ran83,lm88}.
Most of the pulsars in our sample have been extensively observed at a wide
range of frequencies. We use as much information as possible to
locate $\phi_0$ in the polarization profiles with the most important
being RVM fits where robust.

An added complication is that pulsar emission can occur in two orthogonal
modes which have PAs at right angles to each other \cite{mth75,brc75}.
This can distort the swing of PA
with pulse longitude but also provides a 90\degr\ ambiguity in any analysis of
polarized alignment as already noted by Tademaru (1977).
Figure~\ref{all} illustrates the angles in the problem and shows the
orthogonal emission possibilities. It can be seen that
although one would like to associate the measured PA$_0$ directly with
PA$_r$, it may be that PA$_r$ = PA$_0$ + 90\degr\ if the pulsar emission
is in the orthogonal mode.

\section{Motivation and Source selection}
Given the recent results, the time seems right to re-visit the
question, from an observational point of view, of whether pulsar spin
axes and velocity vectors align.  In order to achieve this, we need
proper motion measurements with small errors, high-quality
polarization data with absolute position angles, and accurate rotation
measures (RM) to remove Faraday rotation from the measured
polarization PAs.  Our
intention was to make high-quality polarization observations on a
number of pulsars for which good proper motion data exist.

There were three main criteria for source selection.  First, the
chosen pulsars had to have accurate proper motions, with an angle on
the sky known to better than 20\degr.  Secondly, the pulsars had to be
relatively young so that the velocity vector is an accurate
representation of their birth velocity vector (i.e. had not had time to
be significantly contaminated by acceleration in the Galactic
potential). Finally, we concentrate here on southern pulsars with
declinations less than +25\degr; a future study of northern pulsars is
planned. We chose a total of 25 pulsars which met most of these
criteria.
We have included PSR~J1709$-$4429 in our sample even though it does not
have a measured proper motion. Dodson \& Golap (2002)\nocite{dg02} and 
Bock \& Gvaramadze (2002)\nocite{bg02} have proposed 
that the pulsar is associated with the supernova remnant
G343.1$-$2.3 and claim that the proper motion vector should thus be
in the range 150\degr\ to 170\degr. Although the claimed association
has been disputed \cite{njk96,fs97}, Ng \& Romani (2004)\nocite{nr04}
derived a position angle vector of 175\degr\ for the rotation axis
based on observations of an X-ray torus around the pulsar.

Table~\ref{sources} lists the pulsars, none of which are binary or
millisecond pulsars. The first two columns list the
J and B names of the pulsars. Column 3 gives the logarithm of the
pulsar's characteristic age, $\tau_c = P/2\dot{P}$, with $P$ the
pulsar period and $\dot{P}$ the period derivative. Column 4 gives
the distance to the pulsar in kpc.  Columns 5 to 8 give
the proper motions in RA and Dec, the reference and PA$_v$, the
position angle of the velocity vector on the sky measured
counter-clockwise from north.  In all cases the number in brackets indicates
the error on the last digit(s).
There are 10 pulsars in common with the
Deshpande et al. (1999) sample and 5 in common with Anderson \& Lyne
(1983).  In most cases, the errors in the direction of the velocity
are now significantly reduced.

In the following sections we describe the polarization observations
and the procedures required to determine absolute position angles at
the pulsar, including the necessary accurate RM determination.  We
present our results in section 5 and 6 and discuss the implications in
section 7.

\section{Observations}
The observations were carried out using the Parkes radio telescope
from 2004 November 27 to 30. For the first two days we used the H-OH
receiver at a central frequency of 1369~MHz with a bandwidth of
256~MHz. On the last two days we used the 10/50~cm receiver at a
central frequency of 3100~MHz with a bandwidth of 512~MHz.  Both
receivers have orthogonal linear feeds and also have a pulsed
calibration signal which can be injected at a position angle of
45\degr\ to the two feed probes.  A digital correlator
was used which subdivided the bandwidth into 1024 frequency channels
and provided all four Stokes' parameters. We also recorded 1024
phase bins per pulse period for each Stokes' parameter.

The pulsars were observed twice for 20 minutes each time, with a feed
rotation of 90\degr\ between observations. Prior to the observation of
the pulsar a 3-min observation of the pulsed calibration signal was
made.  The data were written to disk in FITS format for subsequent
off-line analysis. Data analysis was carried out using the PSRCHIVE
software package \cite{hvm04}.

\section{Data Analysis and Calibration}

In order to determine the intrinsic positional angle of linear
polarization at the pulsar,
there are a number of factors which need to be taken into account.
Working backwards from the backend system to the pulsar, these are
(i) the phase delay in the paths from the feed to the backend,
(ii) gain differences between the two orthogonal probes of the feed,
(iii) possible cross-talk or leakage between the two probes,
(iv) the orientation of the feed probes with respect to the meridian,
(v) the parallactic angle of the source,
(vi) the rotation measure due to the terrestrial ionosphere and
(vii) the rotation measure due to the interstellar medium.

The first two of these can be removed using standard calibration 
techniques. We observed a pulsed calibration signal,
injected at 45\degr\ to the
two probes so that the signal is 100 per cent polarised in Stokes $U$.
Gain corrections can be made by comparing Stokes $Q$ with Stokes $I$
and phase corrections can be made by comparing Stokes $U$ with Stokes $V$
on a channel by channel basis in the output data.
Cross talk, or leakage terms can be largely eliminated by summing
the two scans made with 90\degr\ feed offsets \cite{joh02}.
As a further check we observed PSR~J1359--6038 on 8 separate occasions
over a range of hour angles. This allowed us to determine impurities
in the feed, which we estimate to be no greater than 2 percent.

The RM must be determined to high accuracy because of
the need to account for the Faraday rotation of the polarized signal.
If the error in the RM is denoted $\delta_{{\rm RM}}$ then the
error $\delta_{{\rm PA}}$ (in radians) in the Faraday-corrected PA,
can be calculated as
\begin{equation}
\delta_{{\rm PA}} = \frac{c^2 \,\,\, \delta_{{\rm RM}}}{\nu^2}
\end{equation}
where $c$ is the speed of light and $\nu$ the observing frequency.  At
1369~MHz, therefore, an RM error of 1~rad~m$^{-2}$ yields a PA error
of 2.8\degr.  It has been noted by a number of authors (e.g. Weisberg
et al. 2004\nocite{wck+04}) that RMs vary with epoch and we thus
measured RMs for all the pulsars in our sample rather than
use previous values which can be decades old.  There are two ways to
determine the RM.  The first is to measure the PA rotation across the
256~MHz bandwidth at 1369~MHz.  The second method is to determine the
differences in PA obtained at 1369 and 3100~MHz. The latter gives more
accurate results in principle but profile shapes and 
the intrinsic PA can sometimes be frequency-dependent \cite{kj05}.
For all the pulsars in our
sample for which we have dual-frequency observations, we measured the
RM using the latter technique. For pulsars which we only observed at
1369~MHz, we measured the RM across the band. 

The contributions to the
measured values of RM arising in the ionosphere have been estimated
by integrating a time-dependent model of the ionospheric electron
density and geomagnetic field through the ionosphere along the sight
line between the telescope and the pulsar using code provided by JL Han.
Values of the ionospheric
RM range from 0 to $-$2~rad~m$^{-2}$.
We have subtracted the ionospheric RMs from the measured values
to leave only the interstellar components of the RM. 

As a final check of the entire calibration technique, we also carried
out short observations of PSR~J1359--6038 at 1384~MHz with the
Australia Telescope Compact Array (ATCA), a 6 element interferometer
with a correlator which is capable of (crude) pulsar binning.  We
chose PSR~J1359--6038 because it has a flat PA swing across the pulse
and is not therefore affected by time averaging. We determined the
absolute PA of the radiation using the standard package MIRIAD.  The
PA at the ATCA was the same as that measured at Parkes within the
errors, after the RM was taken into account.  This gives us confidence
that our calibration techniques at Parkes are working effectively and
that we have correctly obtained absolute PAs.

\section{Results}
The measured values of interstellar RM are listed in columns 9 to 11 of
table~\ref{sources} as well as previous measurements, along with
the corresponding reference.  The error in RM is typically less than
0.5~rad~m$^{-2}$ for pulsars with dual frequency measurements but
about 10 times larger for the other pulsars. The small error in RM
is crucial for determining absolute position angles.
\begin{figure}
\centerline{\psfig{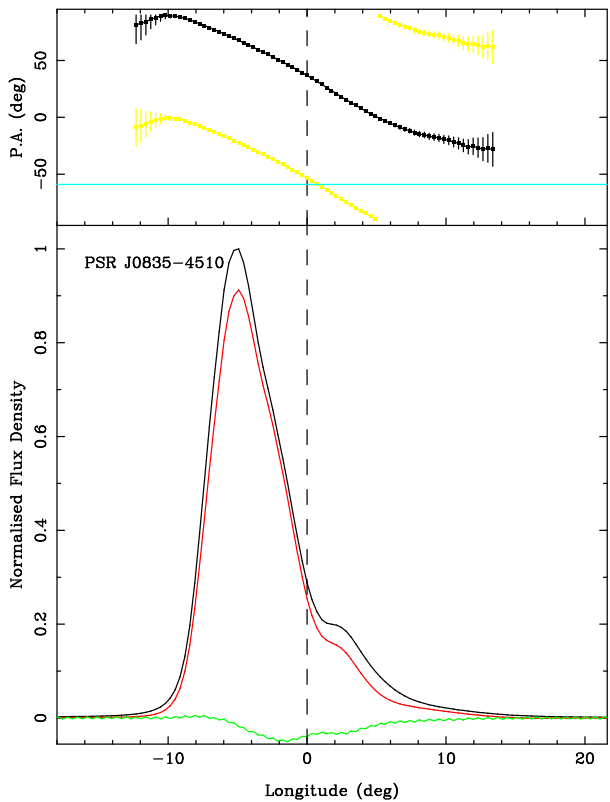}}
\caption{Polarization profile of the Vela pulsar at 1.4~GHz. The bottom
panel shows the total intensity (black, thick line),
linear polarization (grey line) and circular polarization (pale grey,
thin line). The top panel shows the position angle of the polarized radiation
as a function of pulse longitude. Two curves are shown. The first (dark
points) is the measured PA, the second (lighter points) is the
measured PA + 90\degr.
The position angle of the 
velocity vector, PA$_v$, is shown as a solid, horizontal line. 
Pulse longitude 0.0 ($\phi_0$) is the closest approach of 
the line of sight to the magnetic pole as
determined from the rotating vector model fit. The PA at this
longitude is PA$_0$ from Equation 1. See also Fig.~\ref{all} for a
depiction of the relevant angles.}
\label{vela}
\end{figure}

There are a number of
cases in which the RM differs significantly from previous values,
in particular that of the Vela pulsar.
In 1977, the RM of the Vela pulsar was found to be
38~rad~m$^{-2}$ \cite{hmm+77}. In a later paper, Hamilton et al.
(1985)\nocite{hhc85} showed that the RM had increased to
46~rad~m$^{-2}$ and in fact appeared to be linearly increasing since
at least 1970.  As far as we are aware there are no further RM
measurements quoted in the literature.  Our measured value of the RM
(2004 November) shows it to have a value of 31.4~rad~m$^{-2}$,
significantly different to the older measurements.  Clearly, following
an RM increase for at least 15 years (from 1970 to 1985), the RM has
now significantly decreased and is currently below the 1970 value. At
the time where the RM was increasing, the dispersion measure (DM) was
decreasing (see Hamilton et al. 1985\nocite{hhc85}).  We are not aware
of any further DM measurements for Vela but note that DM values depend
strongly on profile alignment between different frequencies and care
should be taken when comparing results from different epochs.

In the rest of this section, we present polarisation measurements of the
individual sources and describe the determination of PA$_0$, the
intrinsic position angle of polarization at $\phi_0$, the longitude of
closest approach of the line-of-sight to the magnetic pole.  Clearly,
our choice of $\phi_0$ is crucial to our arguments and we
justify this choice in detail for each of
the pulsars in the paragraphs below. In particular we have taken into
account evidence such as profile symmetry, profile frequency evolution
and the circular polarization in determining our choice, in a
similar fashion to the morphology work of Rankin (1983) and
Lyne \& Manchester (1988).
The final column of
table~\ref{sources} contains the measured value of PA$_0$ and its
error which is derived from the error in the RM determination for each
pulsar. Apart from the choice of $\phi_0$ itself, additional
observational errors due to signal to noise issues 
contribute only a very small amount. We deal with other sources of
error in Section 6.3 below.

Figures 2 and 3 show the polarization profiles for the 25 pulsars in
the sample. We first present the results for the Vela pulsar as 
an illustrative case study, and then the results for the other 
pulsars in order of increasing Right Ascension.

\noindent
{\bf PSR J0835$-$4510 (Vela; Figure 2):}
Figure~\ref{vela} shows the polarization profile of the Vela pulsar
from our data. This pulsar represents one of the rare cases where
an RVM fit is robust, so we choose $\phi_0$ from the fit.
The RVM fit constrains both $\alpha$ and $\beta$ reasonably well
($-$137\degr\ and 6.5\degr\ for these data using the sign conventions
in Everett \& Weisberg 2001).
The fitted location of the closest approach of the
line of sight to the magnetic pole is at longitude 0.0 in the figure.

\noindent
{\bf PSR J0525+1115 (Figure~3a):}
The profile at 1.4~GHz consists of at least three components. The linear
polarization is low throughout and the circular polarization changes
sign between the first and second component.
Following Weisberg et al. (1999)\nocite{wcl+99}, we identify $\phi_0$ to 
be at the longitude of the circular polarization sign change.

\noindent
{\bf PSR J0601$-$0527 (Figure~3b):}
At this frequency the PA consists of two flat sections, near
$-$42\degr\ in the leading component and at +75\degr\ in the trailing
component. We have placed $\phi_0$ where the sign of circular
polarization changes. At this location there is no linear polarization
and hence we cannot measure PA$_0$. It is possible that $\phi_0$ is
actually somewhat earlier as the leading component is much more
prominent in low frequency observations \cite{gl98}.  Furthermore, at
low frequencies there does appear to be a swing of PA which connects
the two flat regions although this is complicated by an orthogonal
jump.  Absolute PA determination at a lower frequency may resolve this
issue.

\noindent
{\bf PSR J0630$-$2834 (Figure~3c):}
The profile is rather symmetric in this pulsar and we have placed
$\phi_0$ at the centre of symmetry. There is also rather little
evolution of the pulse shape with frequency and it seems likely that
the entire emission originates from a component near $\phi_0$. 
%It is possible to perform an RVM
%fit to the pulsar. The fit is not particularly well constrained but
%determines the $\phi_0$ to be at longitude 8\degr\ due to a kink in
%the PA swing at this location. For reasons given above we consider it
%unlikely that this is the location of the magnetic pole.

\noindent
{\bf PSR J0742$-$2822 (Figure~3d):}
As in the case of the Vela pulsar, the PA swing of this pulsar
permits a robust RVM fit to be obtained and we used the fitted value
of $\phi_0$ to locate the closest approach to the magnetic pole. 
This longitude is also the location of the mid-point of the profile
at the 10 per cent emission level.

\noindent
{\bf PSR J0820$-$1350 (Figure~3e):}
The pulsar shows rather little profile evolution with frequency and
also shows a steep swing of polarization through the middle of the pulse.
We have therefore located $\phi_0$ at the symmetry centre of the
profile. An RVM fit can be attempted for this pulsar; although the fit
is not very good, the fitted $\phi_0$ is coincident with our chosen $\phi_0$.
\begin{figure*}
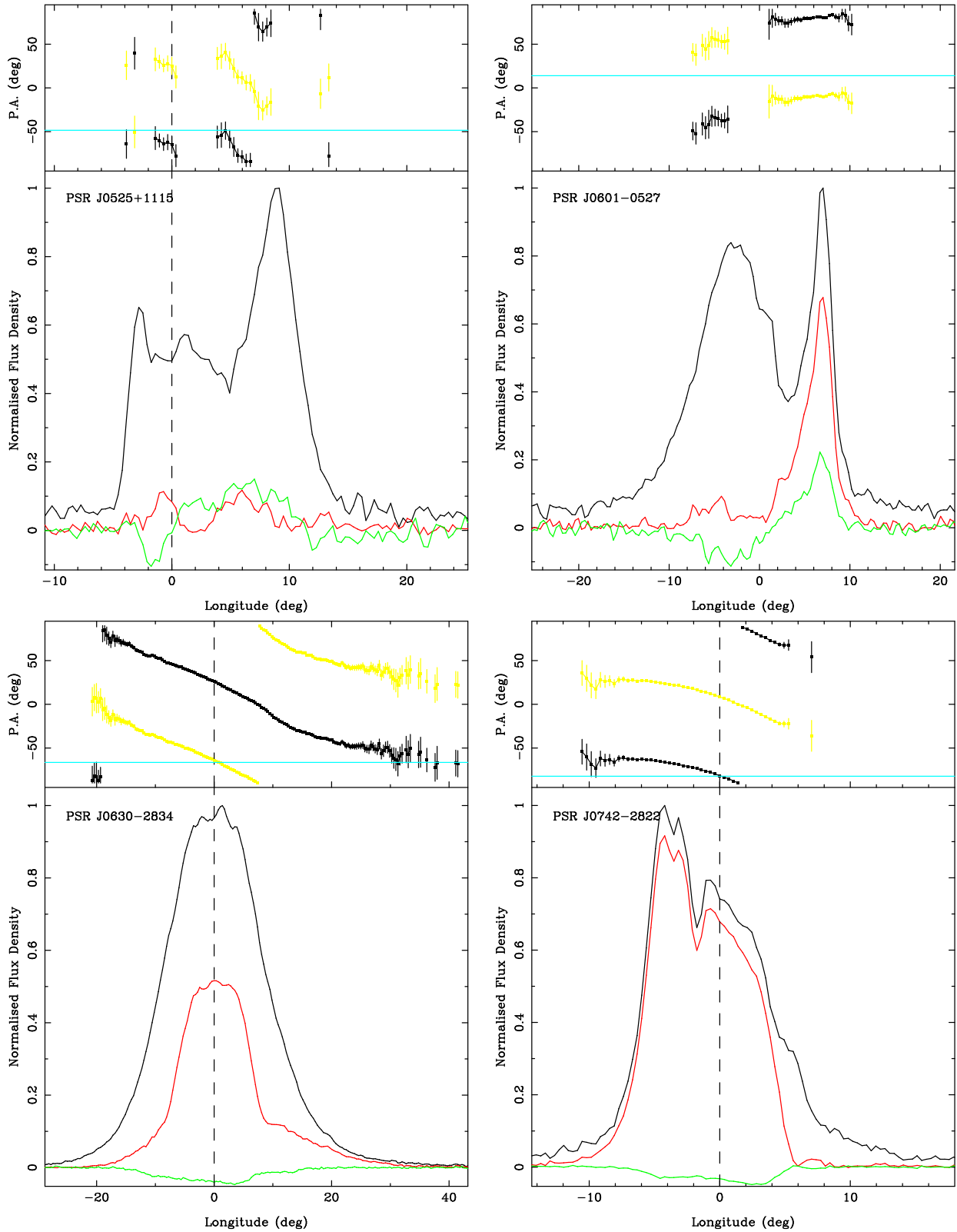

\begin{tabular}{cc}
\psfig{figure=0525_1.ps,angle=0,width=8cm} &
\psfig{figure=0601_1.ps,angle=0,width=8cm} \\
\psfig{figure=0630_1.ps,angle=0,width=8cm} &
\psfig{figure=0742_1.ps,angle=0,width=8cm} \\
\end{tabular}
\caption{(a)-(d). Polarization profiles at 1.4~GHz for
PSR~J0525+1115 (top left),
PSR~J0601$-$0527 (top right), PSR~J0630$-$2834 (bottom left) and
PSR~J0742$-$2822 (bottom right).
See caption for Fig.~\ref{vela} and text for details.
The choice of longitude zero for each pulsar is given in the text and
is marked with a dashed line where an accurate determination can be made.}
\end{figure*}
\addtocounter{figure}{-1}
\begin{figure*}
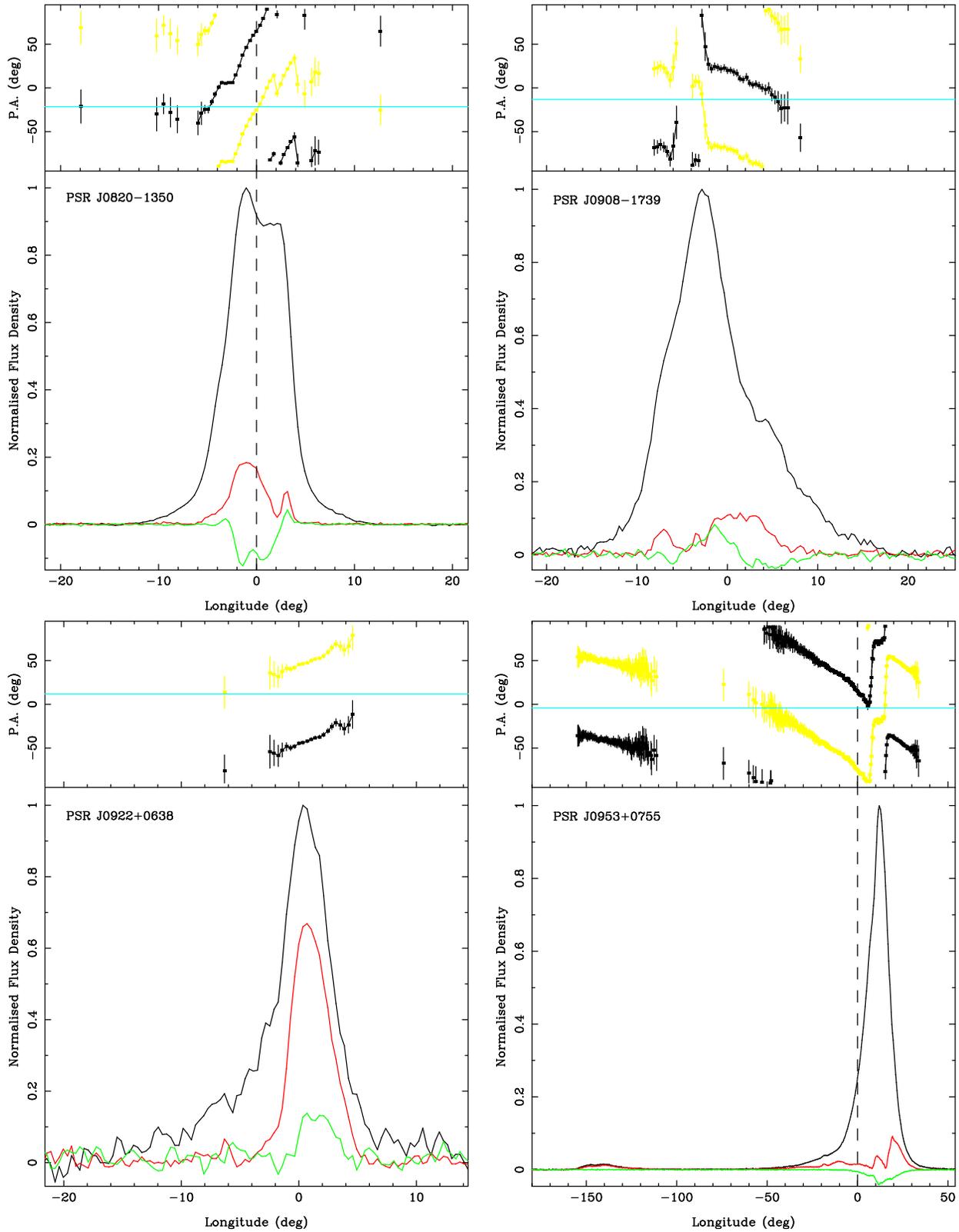

\begin{tabular}{cc}
\psfig{figure=0820_1.ps,angle=0,width=8cm} &
\psfig{figure=0908_1.ps,angle=0,width=8cm} \\
\psfig{figure=0922_1.ps,angle=0,width=8cm} &
\psfig{figure=0953_1.ps,angle=0,width=8cm} \\
\end{tabular}
\caption{(e)-(h). Polarization profiles at 1.4~GHz for
PSR~J0820$-$1350 (top left),
PSR~J0908$-$1739 (top right), PSR~J0922+0638 (bottom left) and
PSR~J0953+0755 (bottom right).
See caption for Fig.~\ref{vela} and text for details.
The choice of longitude zero for each pulsar is given in the text and
is marked with a dashed line where an accurate determination can be made.}
\end{figure*}
\addtocounter{figure}{-1}
\begin{figure*}
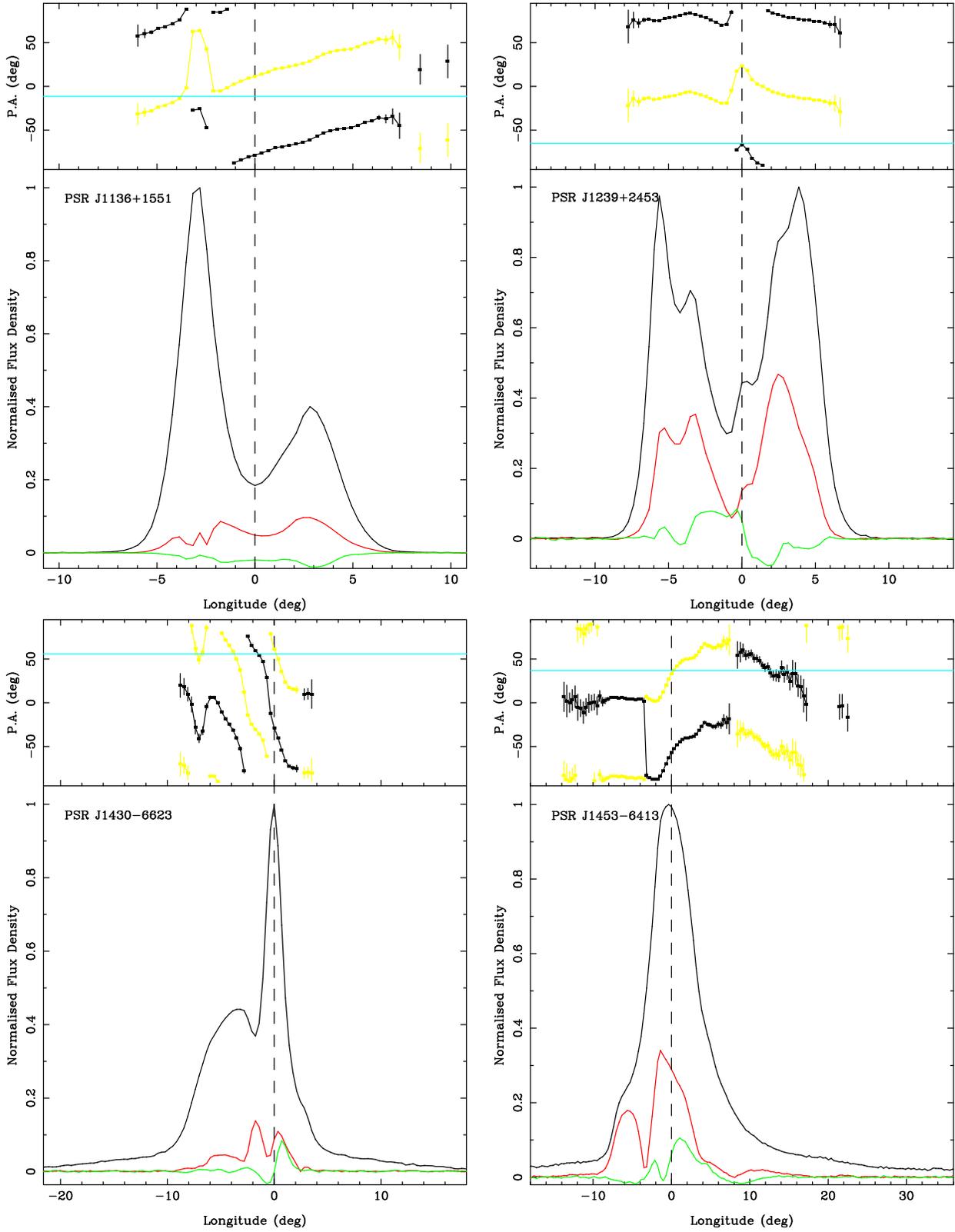

\begin{tabular}{cc}
\psfig{figure=1136_1.ps,angle=0,width=8cm} &
\psfig{figure=1239_1.ps,angle=0,width=8cm} \\
\psfig{figure=1430_2.ps,angle=0,width=8cm} &
\psfig{figure=1453_1.ps,angle=0,width=8cm} \\
\end{tabular}
\caption{(i)-(l). Polarization profiles at 1.4~GHz for
PSR~J1136+1551 (top left),
PSR~J1239+2453 (top right), PSR~J1430$-$6623 (bottom left) and
PSR~J1453$-$6413 (bottom right).
See caption for Fig.~\ref{vela} and text for details.
The choice of longitude zero for each pulsar is given in the text and
is marked with a dashed line where an accurate determination can be made.}
\end{figure*}
\addtocounter{figure}{-1}
\begin{figure*}
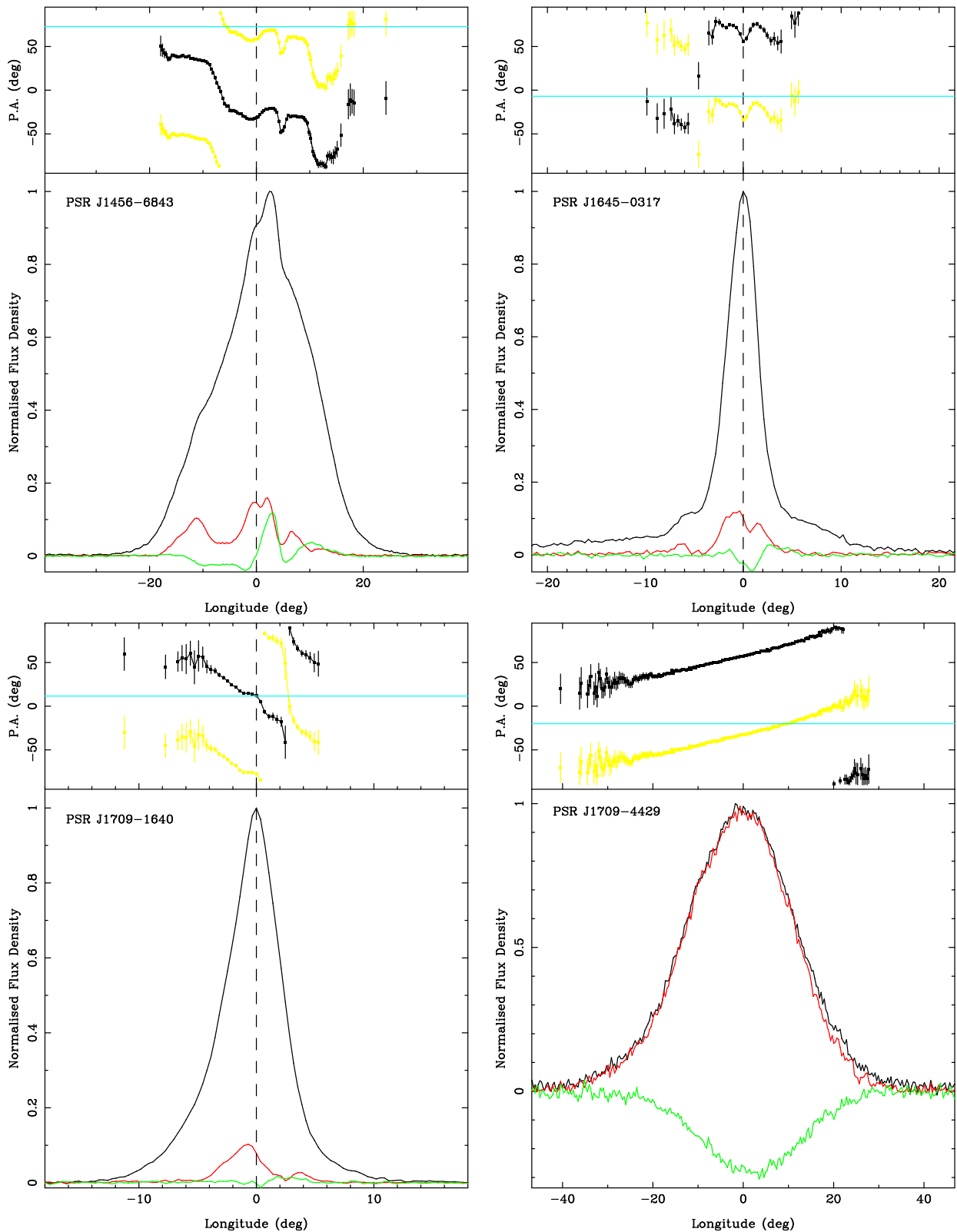

\begin{tabular}{cc}
\psfig{figure=1456_1.ps,angle=0,width=8cm} &
\psfig{figure=1645_1.ps,angle=0,width=8cm} \\
\psfig{figure=1709_1.ps,angle=0,width=8cm} &
\psfig{figure=1706_1.ps,angle=0,width=8cm} \\
\end{tabular}
\caption{(m)-(p). Polarization profiles at 1.4~GHz for
PSR~J1456$-$6843 (top left),
PSR~J1645$-$0317 (top right), PSR~J1709$-$1640 (bottom left) and
PSR~J1709$-$4429 (bottom right).
See caption for Fig.~\ref{vela} and text for details.
The choice of longitude zero for each pulsar is given in the text and
is marked with a dashed line where an accurate determination can be made.}
\end{figure*}
\noindent
\addtocounter{figure}{-1}
\begin{figure*}
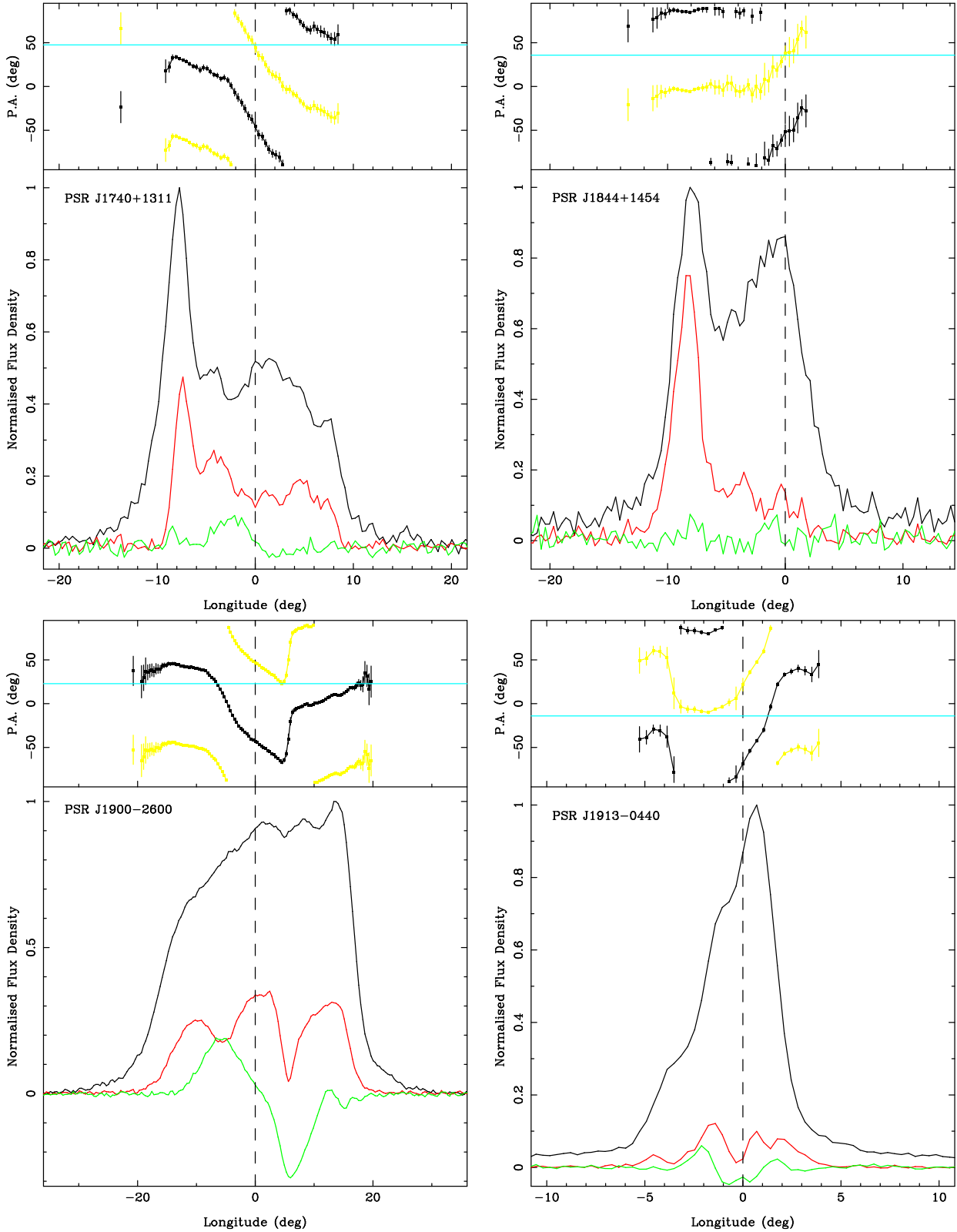

\begin{tabular}{cc}
\psfig{figure=1740_1.ps,angle=0,width=8cm} &
\psfig{figure=1844_1.ps,angle=0,width=8cm} \\
\psfig{figure=1900_1.ps,angle=0,width=8cm} &
\psfig{figure=1913_1.ps,angle=0,width=8cm} \\
\end{tabular}
\caption{(q)-(t). Polarization profiles at 1.4~GHz for
PSR~J1740+1311 (top left),
PSR~J1844+1454 (top right), PSR~J1900$-$2600 (bottom left) and
PSR~J1913$-$0440 (bottom right).
See caption for Fig.~\ref{vela} and text for details.
The choice of longitude zero for each pulsar is given in the text and
is marked with a dashed line where an accurate determination can be made.}
\end{figure*}
\addtocounter{figure}{-1}
\begin{figure*}
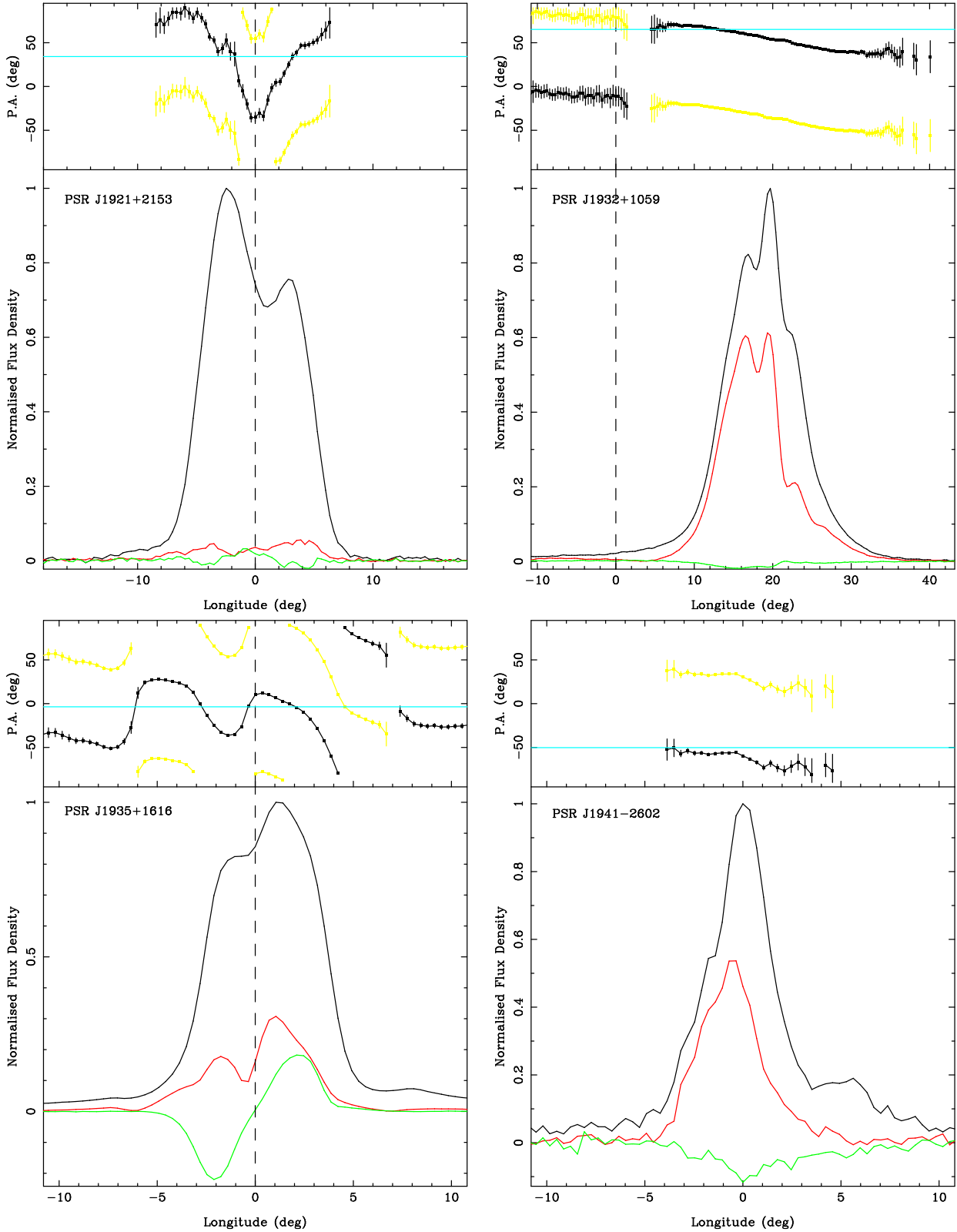

\begin{tabular}{cc}
\psfig{figure=1921_1b.ps,angle=0,width=8cm} &
\psfig{figure=1932_1.ps,angle=0,width=8cm} \\
\psfig{figure=1935_1.ps,angle=0,width=8cm} &
\psfig{figure=1941_1.ps,angle=0,width=8cm} \\
\end{tabular}
\caption{(u)-(x). Polarization profiles at 1.4~GHz for
PSR~J1921+2153 (top left),
PSR~J1932+1059 (top right), PSR~J1935+1616 (bottom left)
and PSR~J1941$-$2602 (bottom right).
See caption for Fig.~\ref{vela} and text for details.
The choice of longitude zero for each pulsar is given in the text and
is marked with a dashed line where an accurate determination can be made.}
\end{figure*}
\noindent

{\bf PSR J0908$-$1739 (Figure~3f):}
This pulsar is classified as a partial cone (likely the leading edge)
by Lyne \& Manchester (1988)\nocite{lm88}. There is virtually no
frequency evolution of the profile between 0.4 and 3.1~GHz lending
support to this idea. In Figure~3f we have simply labelled the profile
midpoint as longitude zero, although its location is likely to be at
significantly later longitudes.   We do not assign $phi_0$ or
$PA_0$ for this pulsar.

\noindent
{\bf PSR J0922+0638 (Figure~3g):}
This pulsar is classified as a partial cone (likely the trailing part)
by Lyne \& Manchester (1988)\nocite{lm88}. Weisberg et
al. (1999)\nocite{wcl+99} provide a comprehensive review of the many
observations of this pulsar.  It is likely that $\phi_0$ is located
somewhat earlier than the centre of emission, where the polarized
emission is too low to determine PA$_0$.

\noindent
{\bf PSR J0953+0755 (Figure~3h):}
The profile consists of a main and interpulse and has (polarized)
emission over virtually the whole 360\degr\ of longitude. RVM fitting
has been attempted by a number of authors as summarised in Everett \&
Weisberg (2001)\nocite{ew01}.  The fit given by those authors shows a
preference for a single-pole interpretation of the polarization PA
swing in this pulsar, unlike other authors who prefer an orthogonal
rotator (e.g. Blaskiewicz, Cordes \& Wasserman 1991)\nocite{bcw91}.
As Everett \& Weisberg (2001)\nocite{ew01} appear to have the best
data, we have used their fit to establish the location of $\phi_0$ at
a longitude which leads the peak of the main pulse by 12\degr.  

\noindent
{\bf PSR J1136+1551 (Figure~3i):}
The pulsar appears to have a classic double-cone profile
and we locate $\phi_0$ at the midpoint of the profile.

\noindent
{\bf PSR J1239+2453 (Figure~3j):}
This pulsar is a classic five-component pulsar. It is likely that the
core component is located close to the centre of the profile flanked
by the inner and outer cones. This interpretation is backed up by the
fact that at high frequencies the core component is less prominent and
the swing of the handedness of circular polarization coincides with
$\phi_0$.  The PA swing in this pulsar does not
conform to the RVM so no value of $\phi_0$ can be obtained from an RVM
fit.  

\noindent
{\bf PSR J1430$-$6626 (Figure~3k):}
This pulsar has an extremely complex PA curve at 1.4~GHz and appears
to swing through 270\degr, not allowed in the RVM model.  There is
rather little profile evolution between 0.4 and 1.4~GHz but the
3.1 GHz observations show a strong decline in the flux density of the
leading component. At the peak of the trailing component, the sign of
circular polarization changes from negative to positive.  It seems
likely that $\phi_0$ is close to the peak of this
component, so we have used this location to estimate PA$_0$.

\noindent
{\bf PSR J1453$-$6413 (Figure~3l):}
The profile of this pulsar evolves strongly with frequency. At high
frequencies, the component leading the main peak dominates the
profile and therefore seems likely to be a conal component. The prominent
peak at 1.4~GHz is likely to be the core component and 
this component also shows a steep swing of PA.
We therefore assign $\phi_0$ to the peak of this
component which also corresponds to a local inflexion point in the PA swing,
although a formal fit to the RVM is not possible.

\noindent
{\bf PSR J1456$-$6843 (Figure~3m):} Although there appear to be
multiple components in the profile of this pulsar there is rather
little evolution with frequency across a wide frequency
range. The PA swing appears to be simpler at lower
frequencies \cite{hmak77}, although this may be an artifact of low
time resolution. We have placed $\phi_0$ where the sign of circular
polarization changes; this is almost the centre of symmetry of the
profile.

\noindent
{\bf PSR 1645$-$0317 (Figure~3n):}
The pulsar is classified as a core single at low frequencies
as it shows a simple single-component profile \cite{hmak77}.
At 5~GHz and above, the conal outriders become very prominent \cite{hx97}
and we are therefore confident that the peak of the profile marks $\phi_0$.

\noindent
{\bf PSR J1709$-$1640 (Figure~3o):} This pulsar is classified as a core
single. We locate $\phi_0$ at the peak of the profile, which is also
the location at which the circular polarization changes sign.  We were
unable to estimate a value for the RM for this pulsar with sufficient
accuracy to improve on the previously listed value.

\noindent
{\bf PSR J1709$-$4429 (Figure~3p):}
Unfortunately, polarization observations of this pulsar are not particularly
illuminating as it is difficult to tell whether $\phi_0$ is located
near to the single,
rather broad component or whether it is instead
far out in the pulsar beam (as claimed by
Manchester 1996\nocite{man96} and Crawford et al. 2001\nocite{cmk01}).
We therefore do no assign $\phi_0$ or PA$_0$ for this pulsar.

\noindent
{\bf PSR J1740+1311 (Figure~3q):}
We did not obtain high frequency observations of this pulsar and
used the value of RM given in Weisberg et al. (2004)\nocite{wck+04}.
This pulsar is a 5-component pulsar with the core located close to
the centre of the profile (see the discussion in Weisberg et al. 2004) and
we assign $\phi_0$ to this location.
The PA swing also shows an inflection point at this same
longitude, further strengthening the case that this location marks the
magnetic pole crossing. 

\noindent
{\bf PSR J1844+1454 (Figure~3r):}
This pulsar only has a single component at low frequencies.
At higher frequencies a conal outrider becomes prominent and dominates
the profile above 1.4~GHz. We have located $\phi_0$ at the centre of
the strong low frequency component.

\noindent
{\bf PSR J1900$-$2600 (Figure~3s):}
The profile is complex and shows multiple components. The circular
polarization shows a characteristic swing with a change of handedness
near the centre of symmetry of the profile. This fact, coupled
with observations at lower frequencies \cite{mhma78} make 
it likely that $\phi_0$ is near the centre of the profile. 

\noindent
{\bf PSR J1913$-$0440 (Figure~3t):}
The profile consists of three components at 1.4~GHz. At lower
frequencies the trailing component dominates the profile
\cite{hmak77}, whereas at higher frequencies the leading two
components become more dominant. Given that there is also a steep
swing of PA associated with the trailing component we locate
$\phi_0$ near the peak of this component.  

\noindent
{\bf PSR J1921+2153 (Figure~3u):}
Weisberg et al. (1999)\nocite{wcl+99} describe observations of this
pulsar made at various frequencies.  The pulsar is likely a conal double
with $\phi_0$ near the profile centre.
The PA profile deviates strongly from that expected in the RVM model.

\noindent
{\bf PSR J1932+1059 (Figure~3v):}
This pulsar has emission over a large longitude range and also has
a low-amplitude interpulse; in the figure we show only the main pulse.
For a more sensitive
observation at this frequency see Weisberg et al. (1999)\nocite{wcl+99}.
Many authors have attempted an RVM fit for this pulsar, and these
are discussed by Everett \& Weisberg (2001)\nocite{ew01}. We use their
fitted results to locate $\phi_0$. 

\noindent
{\bf PSR J1935+1616 (Figure~3w):}
This pulsar is classified as a core single at low frequencies.  At
high frequencies, conal outriders appear but the central part of the
profile remains constant (see the discussion in Weisberg et al. 1999).
The PA is rather complicated with a number of orthogonal jumps across
the profile and an RVM fit is not possible.  We therefore take the
location where the circular polarization changes hand to be $\phi_0$.

\noindent
{\bf PSR J1941$-$2602 (Figure~3x):}
The profile of this pulsar also seems likely to be a partial cone.
The PA variation is
rather flat and there is virtually no profile evolution over a wide
frequency range. In particular, the shoulder seen at the trailing edge
of the profile remains constant with frequency. We cannot assign
$\phi_0$ or PA$_0$ for this pulsar.

\section{Discussion}
\begin{table}
\caption{PA$_v$, PA$_0$ and their offset, $|\Psi|$, for our sample of
25 pulsars. The figure in brackets gives the error in the last digit(s).}
\begin{tabular}{llrrrl}
\hline & \vspace{-3mm} \\
\multicolumn{1}{c}{Jname} & \multicolumn{1}{c}{Bname} 
& \multicolumn{1}{c}{PA$_v$} & \multicolumn{1}{c}{PA$_0$}
& \multicolumn{1}{c}{$|\Psi|$} & Fig \\
& & \multicolumn{1}{c}{(deg)} & \multicolumn{1}{c}{(deg)}
& \multicolumn{1}{c}{(deg)}\\
\hline & \vspace{-3mm} \\
\multicolumn{6}{c}{Pulsars with $|\Psi|<$10\degr\ or $|\Psi|>$80\degr}\\
J0630$-$2834 & B0628$-$28 & 294(3)   & 26(2) & 88(4) & 3c \\
J0742$-$2822 & B0740$-$28 & 278(5)   & $-$81.7(1) & 0(5) & 3d \\
J0820$-$1350 & B0818$-$13 & 159(6)   & 65(2) & 86(6) & 3e \\
J0835$-$4510 & B0833$-$45 & 301.0(1) & 36.8(1) & 84.2(2) & 2 \\
J1239+2453   & B1237+25   & 295.0(1) & $-$66(1) & 1(1) & 3j \\
J1430$-$6623 & B1426$-$66 & 236(9)   & $-$28.5(7) & 85(9) & 3k \\
J1453$-$6413 & B1449$-$64 & 217(3)   & $-$56.9(4) & 86(3) & 3l \\
J1709$-$1640 & B1706$-$16 & 192(16)  & 15(2) & 3(16) & 3o \\
J1740+1311   & B1737+13   & 227(6)   & $-$46(4) & 87(7) & 3q \\
J1844+1454   & B1842+14   & 36(15)   & $-$52(2) & 88(15) & 3r \\
\multicolumn{6}{c}{Pulsars with 10\degr$<|\Psi|<$~80\degr}\\
J0525+1115   & B0523+11   & 132(16)  & $-$65(4) & 17(16) & 3a \\
J0953+0755   & B0950+08   & 355.9(2) & 14.9(1) & 19.0(2) & 3h \\
J1136+1551   & B1133+16   & 348.6(1) & $-$78(2) & 67(2) & 3i \\
J1456$-$6843 & B1451$-$68 & 252.7(6) & $-$31.6(6) & 76(1) & 3m \\
J1645$-$0317 & B1642$-$03 & 353(3)   & 56(4) & 63(5) & 3n \\
J1900$-$2600 & B1857$-$26 & 202.8(7) & $-$43(2) & 66(2) & 3s \\
J1913$-$0440 & B1911$-$04 & 166(11)  & $-$68(2) & 54(11) & 3t \\
J1921+2153   & B1919+21   & 34(12)   & $-$35(2) & 69(12) & 3u \\
J1932+1059   & B1929+10   & 65.2(2)  & $-$11.3(1) & 76.5(2) & 3v \\
J1935+1616   & B1933+16   & 176(1)   & 10.1(7) & 14(1) & 3w \\
\multicolumn{6}{c}{Pulsars for which PA$_0$ and hence $\Psi$ cannot be determined}\\
J0601$-$0527 & B0559$-$05 & 194(16)  & & & 3b \\
J0908$-$1739 & B0906$-$17 & 167(2)   & & & 3f \\
J0922+0638   & B0919+06   & 12.0(1)  & & & 3g \\
J1709$-$4429 & B1706$-$44 & 160(10)  & & & 3p \\
J1941$-$2602 & B1937$-$26 & 130(17)  & & & 3x \\
%\multicolumn{6}{c}{Pulsars with large errors on $\Psi$}\\
%J0525+1115   & B0523+11   & 132(16)  & $-$65(4) & 17(16) & 3a \\
%J1709$-$1640 & B1706$-$16 & 192(16)  & 15(2) & 3(16) & 3o \\
%J1917+1353   & B1915+13   & 321(28)  & $-$75(10) & 36(30) & 3u \\
\hline & \vspace{-3mm} \\
\end{tabular}
\label{results}
\end{table}

\begin{figure}
\centerline{\psfig{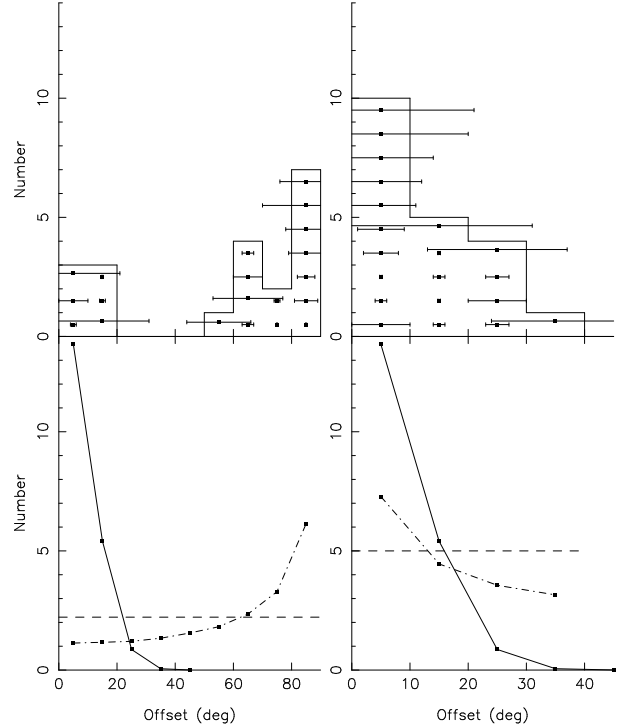}}
\caption{Histograms of observed (top) and simulated (bottom) values of $|\Psi|$.
The left hand panels ignore the possibility of orthogonal mode emission
so that PA$_0$=PA$_r$. The right hand panels include this possibility
which then reduces the maximum offset to 45\degr.
Also included is measurement uncertainties in the data with 
typical value of 10\degr.
Top panels: Observed values of $|\Psi|$ in bins of 10\degr\ for 20 pulsars.
Bottom panels: Simulated histogram for PA$_r$ parallel to PA$_v$ (thick line),
uncorrelated PA$_r$ and PA$_v$ (dashed line) and
PA$_r$ perpendicular to PA$_v$ (dot-dash line).}
\label{histo}
\end{figure}
In Table~\ref{results} we summarise our results. Columns 3 and 4 show
the position angles of the velocity vector, PA$_v$, and of the linear
polarisation, PA$_0$, at $\phi_0$, the closest approach of the
line-of-sight to the magnetic pole (see Equation 1), taken from
Table~\ref{sources}.
Column 5 shows the offset, $|\Psi|$, between PA$_v$ and PA$_0$ where
$\Psi$ is defined as
\begin{equation}
\Psi = {\rm PA}_v - {\rm PA}_0 ; \,\,\,\,\, -90\degr \leq \Psi \leq 90\degr
\end{equation}
We force $\Psi$ to lie between $-$90\degr\ and 90\degr by rotating
PA$_0$ by $\pm$180\degr\ as necessary (note that any polarized PA has
by nature a 180\degr\ ambiguity).
The error in $\Psi$ comes from a quadrature sum of the errors in
PA$_v$ and PA$_0$.

The table is subdivided into three sections.  The top section lists
those pulsars for which we find a good relationship between PA$_0$ and
PA$_v$ defined as where $|\Psi|$ is less than 10\degr\ or greater than
80\degr. There are 10 pulsars in this category.  The second section
lists the 10 pulsars for which there is no clear relation between
PA$_0$ and PA$_v$ i.e. where $|\Psi|$ lies between 10 and
80\degr. Section~3 lists the 5 pulsars where we can not make a judgement
either because there is either no emission at $\phi_0$ or
the PA swing is too complicated to determine PA$_0$.

\subsection{Statistics}
The top left panel of Figure~\ref{histo} shows the observed distribution
of $|\Psi|$ as a histogram for the 20 pulsars for which we have
measured $|\Psi|$.
If we assume that PA$_v$ and PA$_0$ are unrelated then the
expectation is that the offsets between them should be randomly
distributed in the range 0\degr\ to 90\degr. This is shown by the
dashed line on the bottom left panel of Fig.~\ref{histo}.  The
probability of having 7 objects in the top bin, as observed, is very
small, $2.6\times 10^{-4}$.  A Kolmogorov-Smirnov (KS) test rules
out a random distribution at the 94 per cent confidence level.
Clearly our measurements show that PA$_v$
and PA$_0$ must be related in some fashion.  Consider now some
possible types of relationship between the observables PA$_v$ and
PA$_0$; namely that they are either parallel or perpendicular.  To
start, we ignore the confounding factor of the orthogonal emission so
that PA$_0$=PA$_r$ (see Fig.~\ref{all}).

The case where PA$_v$ is parallel to PA$_r$ is a simple one - all the
pulsars should fall in the first bin of the histogram. However,
measurement errors will broaden this distribution as shown by the
solid line on the bottom left panel of Fig.~\ref{histo}.  This also
does not represent the data; the KS test rules out that these distributions
are the same at greater than the 99.9 per cent confidence limit.

The case where PA$_v$ is perpendicular to PA$_r$ is slightly more
complex.  In this case, the projection effects from the true 3-D
vector to the 2-D sky vector will cause the (observed) offset between
PA$_v$ and PA$_0$ to spread away from 90\degr.  This is shown by the
dash-dot line on the bottom left panel of Fig.~\ref{histo} The result
looks remarkably similar to the real data apart from the excess in the
first two bins. The KS test here does {\it not} rule out the possibility
that the distributions are the same.

However, the statistics do not take into account the possible presence
of orthogonal mode emission which complicates the underlying
relationship between PA$_r$ and PA$_0$, as shown in Fig.~\ref{all}.
Either PA$_r$ is parallel to PA$_0$ if the pulsar is emitting in the
`normal' mode or PA$_r$ is perpendicular to PA$_0$ if the pulsar
emission is in the `orthogonal mode'.  Simultaneously allowing for
either possibility serves to reduce the maximum offset between PA$_0$
and PA$_v$ to only 45\degr. The right hand panel of Fig.~\ref{histo}
shows the data and the simulations for such a scenario. The
observed histogram appears somewhat broader than the simulation but
this may be caused by the possible sources of contamination discussed
below. In any case, the results show that we can confidently rule
out that PA$_r$ and PA$_v$ are unrelated (dashed line) at the 98 per
cent confidence limit. The KS test also rules out the case where
PA$_r$ is perpendicular to PA$_v$ (dash-dot line) at greater than the
99 per cent confidence limit but does {\it not} rule out that
the case where PA$_r$ is parallel to PA$_v$ (solid line) is the same
as the observed distribution.

\subsection{Information from outside the radio-band}
In the previous section we have demonstrated that our radio
observations provide clear evidence for a relationship
between the spin and velocity vectors of pulsars. The remaining
uncertainty as to whether parallel and orthogonal vectors occur or
whether the orthogonal PA values are caused by orthogonal emission
modes, can be decided by using information obtained outside the
radio window. We consider these additional pieces of information
in turn.

\subsubsection{The Vela pulsar}
In Section 1 we noted that, in the Vela pulsar, the axis of the X-ray
torus (and hence the pulsar's rotation axis) had a position angle on
the sky of 130\degr\ and 310\degr\ and that this compared well with
the proper motion position angle of 301\degr. In order to compare this
with the {\it polarization} position angle, Helfand et al. (2001) used
the results of Hamilton et al. (1977)\nocite{hmm+77} who measured the
polarization PA {\it at the peak} of the total intensity and obtained
an angle of 64\degr\ (or 244\degr). This angle is about 60\degr\
different from the position angle of the torus and proper motion.
Helfand et al. (2001) then make the bold assertion that as the angle
is ``almost" 90\degr, the emission from Vela must be predominantly in
a mode polarized orthogonally to the magnetic field lines.
Radhakrishnan \& Deshpande (2001)\nocite{rd01} discussed the
implications of the Vela pulsar's orthogonal mode emission in the
context of the physical model of Luo \& Melrose (1995)\nocite{lm95}.

There are two issues here. First, 60\degr\ is not that close to
90\degr.  Secondly, the peak emission of the profile almost certainly
does not coincide with $\phi_0$.  As described above, this is best
measured at the point of greatest swing of the PA curve which, in the
Vela pulsar, occurs significantly later than the pulse peak.  In our
data, the PA at this location appears to be almost exactly 90\degr\
offset from PA$_v$. This provides the observational evidence that the
hypothesis by Helfand et al. (2001) and Radhakrishnan \& Deshpande
(2001) is correct and it is likely 
that the emission in Vela is perpendicular to
the magnetic field direction and that PA$_v$ and the rotation axis
are in fact aligned.

\subsubsection{The Crab pulsar}
Two pieces of evidence from the Crab pulsar also point in this direction.
The Crab has PA$_v$ of 292\degr\ \cite{cm99}, and interpretation of the
X-ray torus indicates that PA$_r$ in the Crab is 124\degr\ or
304\degr\ \cite{nr04} meaning that PA$_r$ is parallel to PA$_v$ within
the measurement uncertainties.

Secondly, while the radio polarization of this pulsar is very complex,
polarization measurements in the optical and UV show a much simpler
picture. The main and inter pulses
show PA variations which are consistent with a RVM and
both have a PA$_0$ near 124\degr \cite{sjdp88,gdb+96},
very similar to the PA of the X-ray torus. In the radio, at low
frequencies the PA of the main and interpulses is 120\degr, whereas
at higher frequencies the PA of the interpulse is 30\degr\ \cite{mh99}.
Although it is unclear where the main and interpulse emission actually
arises (there are competing models for polar cap and light cylinder
emission), it is interesting that the polarization PAs
match both PA$_v$ and PA$_r$.

\subsubsection{Other pulsars}
PSR~B0656+14 has PA$_v$ of 93\degr\ \cite{btgg03}, and recent optical
polarization measurements by Kern et al. (2003)\nocite{kmmh03} match
well with the RVM fit of Everett \& Weisberg (2001)\nocite{ew01}
allowing a determination of PA$_0$ of 98\degr. As we do not have to
worry about orthogonal mode emission at optical wavelengths, then PA$_r$ is
parallel to PA$_v$. For the young pulsar PSR~B0540$-$69 in the Large
Magellanic Cloud, Serafimovich et al. (2004)\nocite{ssls04} have shown
(weak) evidence that the velocity vector and the jet (and hence
rotation axis) align.
Finally, other pulsars in the sample of Ng \& Romani (2004)\nocite{nr04},
PSRs J0538+2817, B1951+32 and B1706--44 also show evidence  
of tori which also indicate that PA$_r$ is parallel to PA$_v$.

\subsection{Possible contaminants}
We are aware of a number of
effects that may influence the observed PA$_v$ and PA$_0$
values, and which may serve to broaden the observed distribution of $\Psi$.
Apart from measurement uncertainties and
orthogonal mode emission already discussed, we can identify four
possible sources of contamination.
These are (i) propagation of the
radiation through the pulsar magnetosphere, (ii) the effects of
differential galactic rotation, (iii) the effects of the galactic
potential on the pulsar's peculiar motion and (iv) the motion of the
pre-supernova progenitor star. We will deal with each of these in
turn.

\subsubsection{Magnetospheric propagation}
Blaskiewicz et al. (1991)\nocite{bcw91} considered the effects of
retardation and aberration of radiation travelling through the pulsar
magnetosphere. In brief their argument implies that the position angle
sweep {\it lags} the profile by an amount which depends on the
altitude at which the emission is radiated. A consequence of this is
that if $\phi_0$ is determined from the RVM fit, it should lag the
$\phi_0$ determined from the profile characteristics.  Furthermore,
under the paradigm that different frequencies are emitted from
different heights in the magnetosphere the effect should be frequency
dependent.
We have estimated the size of the effect by computing an
emission height at our observing frequency using the prescription in
Kijak \& Gil (2003)\nocite{kg03} and then deriving the lag between the
PA curve and the intensity profile. We find that
in the majority of cases this effect is less than 3\degr, but for
the Vela pulsar and PSR~J0742$-$2822 the lag is
13.5 and 8.5\degr\ respectively.
As we have demonstrated, other pieces of evidence suggest
that for Vela our determination of $\phi_0$ is correct and, in any case,
for Vela we have directly measured $\phi_0$ from the RVM and hence
it does not depend on the aberration value.
For PSR~J0742$-$2822, Karastergiou \& Johnston (2005)\nocite{kj05}
have shown that the lag must be significantly smaller than that given
in Kijak \& Gil (2003)\nocite{kg03}.
As the derivation of the emission heights of pulsars is highly model
dependent we have not included these values in our error budget.

\subsubsection{Differential galactic rotation}
The proper motion vector of the pulsar {\it at birth} is what we
really want to compare with PA$_r$, since the latter quantity will
remain fixed in the absence of further torques.  Prior to birth, the
pulsar's progenitor star was orbiting the Galactic Centre with a
velocity which is location dependent.  The difference between this
velocity and the orbital velocity at the Sun's location is the
differential galactic rotation component. The measurement of the
proper motion is made with respect to the solar system barycentre and
therefore includes the effects of differential galactic rotation on
the measured value.  Since birth, the pulsar has moved in the Galactic
potential which has therefore changed its velocity and proper motion
vector.  In order to determine the rotation of the proper motion
vector over the pulsar's lifetime, it is necessary to integrate
equations of motion in the Galactic potential, using as boundary
conditions the pulsar's age, distance, and three-dimensional velocity.
The pulsar's characteristic age, $\tau_c$, listed in Table~1 gives a
useful indication of the true age but may be in error by a factor of 2
and possibly significantly more for young pulsars [e.g. in Vela
\cite{lpgc96} and PSR~J0538+2817 \cite{klh+03}].  The pulsar distance
can be determined from the parallax (where known) or by converting the
dispersion measure using the model of Cordes \& Lazio
(2005)\nocite{cl05} which has an uncertainty for any given object of
up to 50 per cent.  The observed proper motion gives two components of
the velocity, but the radial component is completely unknown although
one can justify that it must lie in the range $\pm$1000~kms$^{-1}$.  A
possible handle on the radial velocity can be obtained in the
following way. If we assume that the pulsar was born in the Galactic
plane and we know its current location and age today, we can adjust
the radial velocity so that extrapolating backwards in time forces a
birth place in the Galactic plane.  Such a technique has been used
relatively successfully in the case of the relativistic binary pulsars
\cite{wkk00,wkh04}.  However, given the large uncertainties in the age and
distance to the pulsars, we do not feel this approach is appropriate
here.

Estimating the differential galactic rotation effect on PA$_v$ at
birth is therefore difficult as we do not know the pulsar's birth
location. We can make some general statements, however, using the
potential model of Kuijken \& Gilmore (1989)\nocite{kg89} which yields
a rotation speed of the Sun of 221.4~kms$^{-1}$.  For young pulsars,
born within $\sim$1.5~kpc of the Sun, the differential galactic
rotation contributes no more than 10-20~kms$^{-1}$ to the overall velocity.
We have simulated this effect on our sample and find that it will
contribute no more than 5\degr\ to the error budget except for very
slow moving pulsars. For the most distant pulsar in our sample,
PSR J1935+1616 (for which $\Psi$=14\degr), 
at a distance in excess of $\sim$5~kpc, the effect could contribute as
much as 20\degr to the error bar.
It is therefore important to ensure that the sample
contains young, nearby pulsars to minimise this effect.

\subsubsection{Acceleration in the galactic gravitational potential}
The effect of the galactic potential on the pulsar's peculiar motion
depends critically on the unknown radial velocity. Again, we can make
only general statements about the magnitude of this effect.
For pulsars born in the vicinity of the Sun, and moving with a velocity
of $\sim$300~kms$^{-1}$, the time-scale for the proper motion vector
to have significantly changed is 15~Myr.
We note that 4 of the 5 pulsars with characteristic ages 
of 3~Myr or less have either $|\Psi|<$10\degr\ or $|\Psi|>$80\degr.
Conversely, of the 6 pulsars with ages greater than 10~Myr,
only one (PSR J1239+2453) has close match between PA$_0$ and PA$_v$.
For the 9 remaining pulsars, with intermediate ages, 5 show a correlation
between PA$_0$ and PA$_v$.

\subsubsection{Motion of the progenitor star}
Finally, an important effect can occur when the pre-supernova star
already has a peculiar velocity. Many high-mass stars are in
binary pairs often with short orbital periods. The pre-supernova star
can therefore already have a velocity of $\sim$100~kms$^{-1}$ or more.
If the imparted kick velocity is also only of this order, then the
resultant velocity vector will not align with the rotation axis.
This scenario was already discussed by Deshpande et al. (1999)
who considered it a major reason for their apparent lack of correlation.

\subsection{Summary}
In summary, we have shown that the case for an
alignment of the velocity and spin vectors is very strong. The case
is strengthened by the fact that the youngest pulsars in the sample,
which suffer from the smallest unknown errors, show the best
alignment.  In addition, we know that in
the case of Vela, the observed emission is in an orthogonal mode.
We think it likely therefore that PA$_v$ and PA$_r$ are parallel
in all cases but that many (though not all) pulsars emit predominantly in 
the orthogonal mode. This fits with the values returned from the KS tests
on the histrograms in Fig~\ref{histo} as discussed in section 6.1.

\section{Implications}
We see no compelling reason to revive the old rocket model
of Tademaru \& Harrison (1975)\nocite{th75} even though that model
predicts that PA$_v$ should be parallel to PA$_r$. The problems with that
model still remain; it requires pulsar spins of 1~ms at birth, it has
difficulty explaining the binary neutron star systems,
and polarization data do not favour off-centred dipoles.

In the context of the Spruit \& Phinney (1998)\nocite{sp98} model
the parallel and orthogonal cases arise in very different ways.
In the perpendicular case, the number of impulses must be small and of
short duration. In the parallel case, the impulses must be of long
duration and can be significant in number.
In principle, the two cases lead to different observables. In the
perpendicular case the (initial) spin period and velocity should
be correlated. In the parallel case, short period pulsars should
have a lower mean velocity. In practise, however, distinguishing
these cases observationally is difficult, although we note that
there is some evidence that short period pulsars are slow moving.
Theoretical arguments based on neutrino scattering and parity
violation tend to favour long duration kicks but this might only
be applicable in magnetic fields greater than 10$^{15}$~G \cite{al99},
values which seem unlikely for the radio pulsars discussed here.
More recently, Schmitt, Shovkovy \& Wang (2005) have speculated
that a neutrino propulsion mechanism associated with the presence
of a colour super-conductor in the transverse A phase, produces
both high kick velocities and aligned spin and velocity vectors.
It is clear that there are theoretical possibilities still to
be explored (see also Lai, Chernoff \& Cordes 2001\nocite{lcc01}).

Both Iben \& Tutokov (1996)\nocite{it96} and
Spruit \& Phinney (1998)\nocite{sp98} argue that
pre-supernova stars do not possess sufficient angular momentum
to explain the fast spins of new-born neutron stars. This
leads the latter authors to propose their model where
one or more off-centre natal kicks is the principal source of
young pulsars' short periods.
We wish to point out an important general consequence of such
models which seems to have escaped notice, namely that off-centre
kick(s) at a radius $r$ that significantly change the magnitude 
of the pulsar's spin angular
momentum will probably also change the angular momentum
{\it direction}. This
is clear from the vector nature of the equation governing
changes in angular momentum, $d\vec{L} / dt = r \times \vec{F}$, 
where  $\vec{F}$ is the force applied.
Unless off-centre kick(s) are somehow applied in a fashion that preserves
the original rotational symmetry of the star, the angular momentum vector
will tip significantly.

Observations of the current parameters of the double neutron star
systems allow a determination of the pre-supernova evolution
of the system \cite{wkk00,wkh04,tds05}. In turn, the kick magnitude
and direction imparted to the (second-born) neutron star can
be constrained. In PSRs B1913+16, B1534+12 and J0737$-$3039
the kick direction is significantly different from the spin axis of
the pre-supernova star and the orbital angular momentum, with the
inference that aligned kicks are therefore ruled out.
However if, as shown above, the kick
imparts both linear {\it and} angular momentum to the neutron star,
then the spin axis of the newly-born neutron star is not related to
either that of its progenitor, or the orbital angular momentum.
Consequently, the derived constraints on the kick direction
can shed no light on the alignment of spin and velocity vectors.

There is an observational test which can be made. If Spruit \& Phinney
(1998)\nocite{sp98} are correct, then the misalignment between the
spin and orbital angular momenta in the PSR~J0737$-$3039 system should
be different for the A and B pulsars whereas if the kick only imparts
velocity and leaves the pre-supernova spin axis unchanged then the
misalignment angles should be identical.
These measurements should be possible in the near future.
Note that even if the misalignment angles are the same, the spin axes of
the two pulsars will not be parallel at the present epoch 
since their precession rates are different.

\section{Conclusions}
We have observed 25 pulsars at 1369~MHz,  21 of which were
also observed at 3100~MHz. Careful polarization calibration,
including accurate RM determination, has enabled
us to compute the PA of the polarized radiation at the pulsar 
to within an accuracy of 2-3\degr. 
We have used all available information
to deduce $\phi_0$, the longitude of the closest approach to the 
magnetic pole of the star. We then compared the PA of the radiation
at this longitude, which is related to the position angle of the axis
of rotation, with the velocity vector derived from proper motion
measurements.

The combination of precise polarization 
measurements, accurate RMs and improved velocity information makes 
this study much more sensitive to any underlying correlation 
between the spin and velocity vectors. In contrast to previous studies
we restricted our sample to young, nearby pulsars where the effects
of contaminants discussed earlier have little or no impact.

We find a clear relationship between the velocity vector and the
rotation axis. The presence of orthogonal modes in pulsar emission
makes it difficult to determine whether the velocity vector is
parallel to the rotation axis or perpendicular to it.  However,
additional information from the optical and X-ray bands for PSR~B0656+14,
the Crab and Vela pulsars allows us to break this ambiguity: the
velocity vector is parallel to the rotation axis and 
many pulsars emit linear polarisation with PA predominantly perpendicular 
to the magnetic field lines.

We point out that if the kicks impart both linear and angular momentum
to the new-born neutron star, the information about the stellar axis
prior to the supernova explosion is lost. Observations of double
neutron star systems cannot therefore rule out aligned kicks, contrary
to claims in the literature.

\section*{Acknowledgments}
The Australia Telescope is funded by the Commonwealth of 
Australia for operation as a National Facility managed by the CSIRO.
We are grateful to W.~van Straten and A.~Hotan for
software support for this project. We thank J.~Han and R.~Manchester
for providing us with ionospheric RM calculation algorithms and
R.~Edwards and A.~Karastergiou for useful conversations.
SV and JMW  acknowledge financial support from
U.S. National Science Foundation grant AST 0406832.

\bibliography{modrefs,psrrefs,crossrefs}
\bibliographystyle{mn}
\label{lastpage}
\end{document}